\documentclass[final,3p,times]{elsarticle}

\usepackage{amssymb}
\usepackage{textgreek}
\usepackage{amsmath}
\usepackage{hyperref}

\journal{Current Opinion in Chemical Biology}

\begin{document}

\begin{frontmatter}



\title{Protein sequence design with deep generative models}


\author[inst1]{Zachary Wu\fnref{inst4}}
\fntext[inst4]{Present address: Google Deepmind, 6 Pancras Square, London N1C, UK}
\affiliation[inst1]{organization={Division of Chemistry and Chemical Engineering, California Institute of Technology},
            addressline={1200 E California Blvd}, 
            city={Pasadena},
            postcode={91125}, 
            state={CA},
            country={USA}}

\author[inst2]{Kadina E. Johnston}
\author[inst1,inst2]{Frances H. Arnold}
\author[inst3]{Kevin K. Yang\corref{corr}}
\cortext[corr]{Corresponding author}
\ead{yang.kevin@microsoft.com}

\affiliation[inst2]{organization={Division of Biology and Biological Engineering, California Institute of Technology},
            addressline={1200 E California Blvd}, 
            city={Pasadena},
            postcode={91125}, 
            state={CA},
            country={USA}}
            
\affiliation[inst3]{organization={Microsoft Research New England},
            addressline={1 Memorial Drive}, 
            city={Cambridge},
            postcode={02142}, 
            state={MA},
            country={USA}}

\begin{abstract}
Protein engineering seeks to identify protein sequences with optimized properties. When guided by machine learning, protein sequence generation methods can draw on prior knowledge and experimental efforts to improve this process. In this review, we highlight recent applications of machine learning to generate protein sequences, focusing on the emerging field of deep generative methods.
\end{abstract}



\begin{keyword}
Deep learning \sep Generative models \sep Protein engineering
\end{keyword}

\end{frontmatter}


\section{Introduction}
Proteins are the workhorse molecules of natural life, and they are quickly being adapted for human-designed purposes. These macromolecules are encoded as linear chains of amino acids, which then fold into dynamic 3-dimensional structures that accomplish a staggering variety of functions. To improve proteins for human purposes, protein engineers have developed a variety of experimental and computational methods for designing sequences that fold to desired structures or perform desired functions \cite{romero2009exploring, arnold2018directed, huang2016coming, garcia2017computational}. A developing paradigm, machine learning-guided protein engineering, promises to leverage the information obtained from wet-lab experiments with data-driven models to more efficiently find desirable proteins \cite{yang2019machine, mazurenko2019machine, volk2020biosystems} . 

Much of the early work has focused on incorporating discriminative models trained on measured sequence-fitness pairs to guide protein engineering \cite{yang2019machine}. However, methods that can take advantage of unlabeled protein sequences are improving the protein engineering paradigm. These methods rely on the metagenomic sequencing and subsequent deposition in databases such as UniProt \cite{uniprot2019uniprot}, and continued development of these databases are essential for furthering our understanding of biology.

Additionally, while studies incorporating knowledge of protein structure are becoming increasingly powerful \cite{ingraham2019generative, sabban2020ramanet, bepler2019learning, anand2020protein, hie2020leveraging_note}, they are beyond the scope of this review, and we focus on deep generative models of protein sequence. \textit{For further detail on protein structure design, we encourage readers to consult Huang and Ovchinnikov's review in this issue of Current Opinion in Chemical Biology.}

In discriminative modeling, the goal is to learn a mapping from inputs to labels by training on known pairs. In generative modeling, the goal is to learn the underlying data distribution, and a deep generative model is simply a generative model parameterized as a deep neural network. Generative models of proteins perform one or more of three fundamental tasks:

\begin{enumerate}
    \item Representation learning: generative models can learn meaningful representations of protein sequences. 
    \item Generation: generative models can learn to sample protein sequences that have not been observed before. 
    \item Likelihood learning: generative models can learn to assign higher probability to protein sequences that satisfy desired criteria.
\end{enumerate}

\begin{figure}
	\centering
	\includegraphics[scale=0.4]{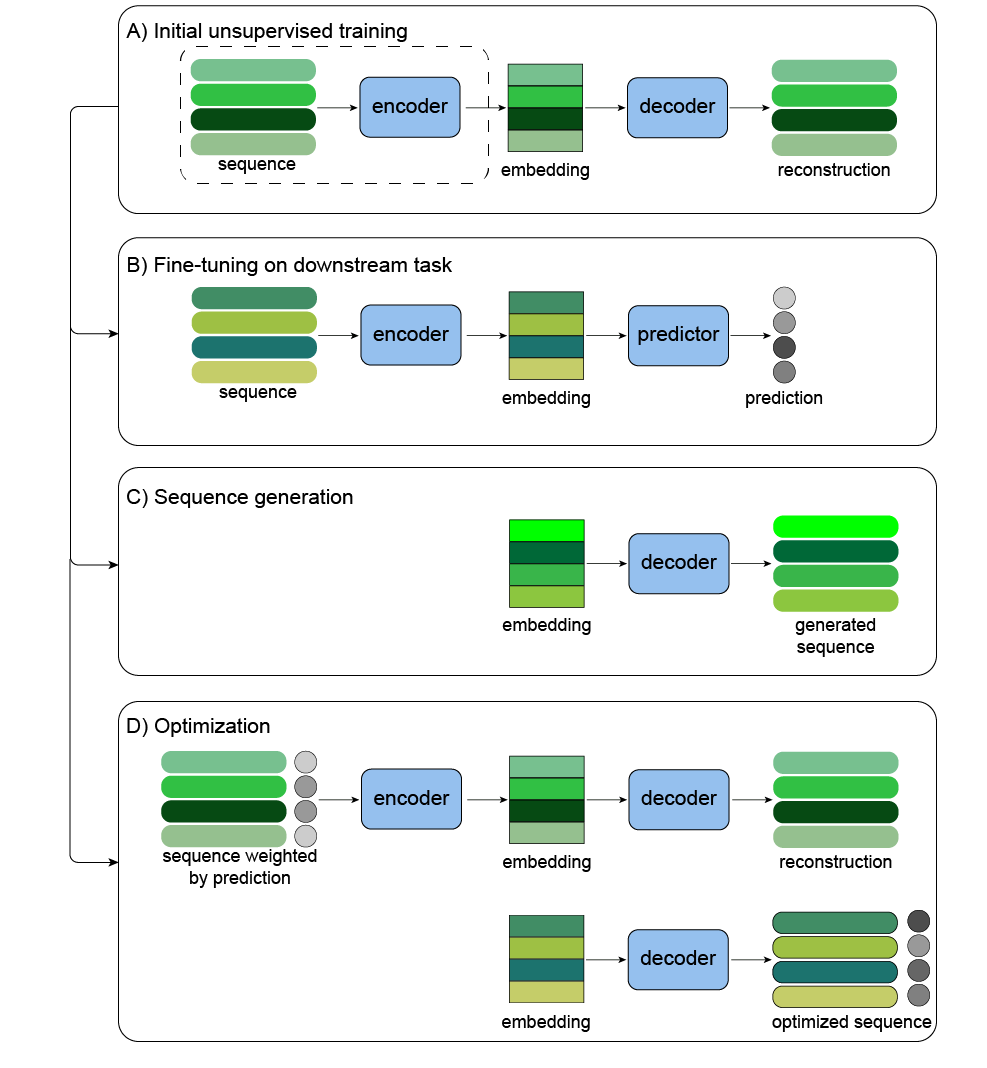}
	\caption{During unsupervised training (A), a generative decoder learns to generate proteins similar to those in the unsupervised training set from embedding vectors. The embeddings can then be used as inputs to a downstream modeling task (B). The decoder can be used to generate new functional sequences (C), or the entire generative model can be tuned to generate functional sequences optimized for a desired property (D).}
	\label{fig:flow}
\end{figure}

In this review, we discuss three applications of deep generative models in protein engineering roughly corresponding to the above tasks: (1) the use of learned protein sequence representations and pretrained models in downstream discriminative learning tasks, an important improvement to an established framework for protein engineering; (2) protein sequence generation using generative models; and (3) optimization by tuning generative models so that higher probability is assigned to sequences with some desirable property. Where possible, these methods are introduced with case studies that have validated generated sequences \textit{in vitro}. Figure~\ref{fig:flow} summarizes these three applications of generative models. Additionally, we provide an overview of common deep generative models for protein sequences, variational autoencoders (VAEs), generative adversarial networks (GANs), and autoregressive models in Appendix A for further background. 

\section{Fine-tuning on downstream tasks}
An established framework for applying machine learning to guide protein engineering is through the training and application of discriminative regression models for specific tasks, which is better reviewed elsewhere \cite{yang2019machine, mazurenko2019machine}. Early examples of this approach were developed by Fox \cite{fox2007improving} and Liao \cite{liao2007engineering} in learning the relationship between enzyme sequence and cyanation or hydrolysis activity, respectively. Briefly, in this approach, sequence-function experimental data are used to train regression models. These models are then used as estimates for the true experimental value, and can be used to search through and identify beneficial sequences \textit{in silico}. 

Learned representations have the potential to be more informative than one-hot encodings of sequence or amino-acid physico-chemical properties. They encode discrete protein sequences in a continuous and compressed latent space, where further optimization can be performed. Ideally, these representations capture contextual information \cite{xu2020deep} that simplifies downstream modeling. However, these representations do not always outperform simpler representations given sufficient training data \cite{shanehsazzadeh2020transfer}.

For example, in BioSeqVAE, the latent representation was learned from 200,000 sequences between 100 and 1000 amino acids in length obtained from SwissProt \cite{costello2019hallucinate}. The authors demonstrate that a simple random forest classifier from scikit-learn \cite{pedregosa2011scikit} can be used to learn the relationship between roughly 60,000 sequences (represented by the outputs of the VAE encoder) and their protein localization and enzyme classification (by Enzyme Commission number) in a downstream fine-tuning task. By optimizing in latent space for either property through the downstream models and decoding this latent representation to a protein sequence, the authors generate examples that have either one or both desired properties. Although the authors did not validate the generated proteins \textit{in vitro}, they did observe sequence homology between their designed sequences and natural sequences with the desired properties.

While the previous study used the output from a pretrained network as a fixed representation, another approach is to fine-tune the generative network for the new task. Autoregressive models are trained to predict the next token in a sequence from the previous tokens (Appendix A). When pretrained on large databases of protein sequence, they have stronger performance than other architectures on a variety of downstream discriminative tasks~\cite{alley2019unified,rives2019biological_note,rao2019evaluating,bepler2019learning}. There are few examples of experimental validation in this space, likely due to the delay in physically verifying computational predictions. However, Biswas and coauthors demonstrated that a double  fine-tuning scheme results in discriminative models that can find improved sequences after training on very few measured sequences~\cite{biswas2020low_note}. First, they train an autoregressive model on sequences in UniRef50~\cite{suzek2015uniref}. They then fine-tune the autoregressive model on evolutionarily-related sequences. Finally, they use the activations from the penultimate layer to represent each position of an input protein sequence in a simple downstream model (in a second round of fine-tuning), showing promising results on two tasks: improving the fluorescence activity of \textit{Aequorea victoria} green fluorescent protein (avGFP) and optimizing TEM-1 {\textbeta}-lactamase. After training on just 24 randomly-selected sequences, this approach consistently outperforms one-hot encodings with 5 to 10 times the hit rate (defined as the fraction of proposed sequences with activity greater than wild type). The authors show that the pre-trained representation separates functional and non-functional sequences, allowing the final discriminator to focus on distinguishing the best sequences from mediocre but functional ones. While the previous work randomly identify the initial set for model training, Wittmann demonstrates an approach which chooses the most informative sets of sequences for further optimized evolution \cite{wittmann2020machine}.

\section{Protein sequence generation}
In addition to improving function predictions in downstream modeling, generative models can also be used to generate novel functional protein sequences. Here, we describe recent successful examples of sequences generated by VAEs, GANs, and autoregressive models.

Hawkins and coauthors generate functional luciferases from two VAE architectures~\cite{hawkins2021generating}: 1) by computing the alignment first and training a VAE (MSA VAE) on the aligned sequences and 2) by introducing an autoregressive component to the decoder to learn the unaligned sequences (AR VAE). Motivated by a similar model used for text generation \cite{semeniuta2017hybrid}, the decoder of the AR VAE contains an up-sampling component, which converts the compressed representation to the length of the output sequence, and an autoregressive component. Both models were trained with roughly 70,000 luciferase sequences (${\sim}360$ residues) and were quite successful: 21/23 and 18/24 variants generated with the MSA VAE and AR VAE (respectively) showed measurable activity. 

The authors of ProteinGAN successfully trained a generative adversarial network to generate functional malate dehydrogenases \cite{repecka2021expanding}. In one of the first published validations of GAN-generated sequences, after training with nearly 17,000 unique sequences (average length: 319), 24\% of 20,000 sequences generated by ProteinGAN display near wild-type level activity, including a variant with 106 mutations to the closest known sequence. Interestingly, although the positional entropy of the final set of sequences closely matched that of the initial input, the generated sequences expand into new structural domains as classified by CATH \cite{sillitoe2019cath}, suggesting structural diversity in the generated results.

Riesselman and coauthors applied autoregressive models to generate single domain antibodies (nanobodies) \cite{riesselman2019accelerating}. As the nanobody's complementarity-determining region is difficult to align due to its high variation, an autoregressive strategy is particularly advantageous because it does not require sequence alignments. With 100,000s of antibody sequences, the authors trained a residual dilated convolutional network over 250,000 updates. While other (recurrent) architectures were tested to capture longer range information, exploding gradients were encountered, as is common in these architectures. After training, the authors generated over 37 million new sequences by sampling amino acids at each new position in the sequence. Further clustering, diversity selection, and removal of motifs that may make expression more challenging (such as glycosylation sites and sulfur residues) enabled the researchers to winnow this number below 200,000, for which experimental results are pending.

Wu \textit{et al.} applied the Transformer encoder-decoder model \cite{vaswani2017attention} to generating signal peptides for industrial enzymes \cite{wu2020signal}. Signal peptides are short (15-30 amino acid) sequences prepended to target protein sequences that signal the transport of the target sequence. After training with 25,000 pairs of target and signal peptide sequences, we generated signal peptide sequences to test \textit{in vitro}, finding that roughly half of the generated sequences resulted in secreted and functional enzymes in \textit{Bacillus subtilis}. 

While this work suffices as early experimentally verified examples, there are many improvements that can be made, such as introducing information upon which to condition generation. Sequences are typically designed \textit{for} a specific task, and task-specific information can be incorporated in the training process \cite{sohn2015learning}. For example, a VAE decoder can be conditioned on the identity of the metal cofactors bound \cite{greener2018design}. After training on 145,000 enzyme examples in MetalPDB \cite{andreini2012metalpdb}, the authors find a higher fraction of desired metal-binding sites observed in generated sequences. Additionally, 11\% of 1000 sequences generated for recreating a removed copper-binding site identified the correct binding amino acid triad. The authors also applied this approach to design specific protein folds, validating their results with Rosetta and molecular dynamics simulations. In ProGen, Madani and co-authors condition an autoregressive sequence model on protein metadata, such as a protein's functional and/or organismal annotation \cite{madani2020progen_note}. While this work does not have functional experimental validation, after training on 280 million sequences and their annotations from various databases, the authors show that computed energies from Rosetta \cite{alford2017rosetta} of the generated sequences are similar to that of natural sequences.

\section{Optimization with Generative Models}

While much of the existing work is designed to generate \textit{valid} sequences, eventually, the protein engineer expects \textit{improved} sequences. An emerging approach to this optimization problem is to optimize with generative models \cite{brookes2019conditioning_note, angermueller2020rmodel_note, amimeur2020designing}. Instead of generating viable examples, this framework trains models to generate \textit{optimized} sequences by placing higher probability on improved sequences (Figure \ref{fig:flow} C). 


One approach to optimization is to bias the data fed to a GAN. Amimeur and coauthors trained Wasserstein GANs \cite{arjovsky2017wasserstein} on 400,000 heavy or light chain sequences from human antibodies to generate regions of 148 amino acids of the respective chain~\cite{amimeur2020designing}. After initial training, by biasing further input data on desired properties (length, size of a negatively-charged region, isoelectronic point, and estimated immunogenicity), the estimated properties of the generated examples shift in the desired direction. While it is not known what fraction of the 100,000 generated constructs is functional from the experimental validation, extensive biophysical characterization of two of the successful designs show promising signs of retaining the designed properties \textit{in vitro}. An alternative study developing a Feedback GAN (FBGAN) framework extends this by iteratively generating sequences from a GAN, scoring them with an oracle, and replacing the lowest-scoring members of the training set with the highest-scoring generated sequences~\cite{gupta2019feedback}. 

Fortunately, this optimization can be enforced algorithmically. The Design by Adaptive Sampling algorithm~\cite{brookes2018design} improves the iterative retraining scheme by using a probabilistic oracle and weighting generated sequences by the probability that they are above the $Q^{\text{th}}$ percentile of scores from the previous iteration. This allows the optimization to become adaptively more stringent and guarantees convergence under some conditions. The authors validate this approach on synthetic ground truth data by training models (of a different type) on real biological data. They then show that generated sequences outperform traditional evolutionary methods (and the previously mentioned FBGAN) when restricted to a budget of 10,000 sequences. The current iteration of this work, Conditioning by Adaptive Sampling (CbAS), improves this approach by avoiding regions too far from the training set for the oracle \cite{brookes2019conditioning}, while other approaches focus the oracle as design moves between regions of sequence space \cite{fannjiang2020autofocused} or emphasize sequence diversity in generations \cite{linder2020generative_note}.

Another approach \cite{angermueller2020rmodel_note} for model-based optimization has roots in reinforcement learning (RL) \cite{sutton2018reinforcement}. The RL framework is typically applied when a decision maker is asked to choose an action that is available given the current state. From this action, the state changes through the transition function with some reward. When a given state and action are independent of all previous states and actions (the Markov property), the system can be modeled with Markov decision processes. This requirement is satisfied by interpreting the protein sequence generation as a process where the sequence is generated from left to right. At each time step, we begin with the sequence as generated to that point (the current state), the select the next amino acid (the action), and add that amino acid to the sequence (the transition function). The reward remains 0 until generation is complete, and the final reward is the fitness measurement for the generated sequence. The action (the next amino acid) is decided by a policy network, which is trained to output a probability over all available actions based on the sequence thus far and the expected future reward. Notably, the transition function is simple (adding an amino acid), so only the reward function needs to be approximated. 

The major challenge under the RL framework is then determining the expected reward. To tackle this issue, Angermueller and coauthors use a panel of machine learning models, each learning a surrogate fitness function $\hat{f_j}$ based on available data from each round of experimentation \cite{angermueller2020rmodel_note}. The subset of models from this panel that pass some threshold accuracy (as empirically evaluated by cross validation) is selected for use in estimating the reward, and the policy network is then updated based on the estimated reward. Thus, this algorithm enables using a panel of models to potentially capture various aspects of the fitness landscape, but only uses the models that have sufficient accuracy to update the policy network. The authors also incorporate a diversity metric by including a term in the expected reward for a sequence that counts the number of similar sequences previously explored. The authors applied this framework to various biologically motivated synthetic datasets, including an antimicrobial peptide (8 - 75 amino acids) dataset as simulated with random forests. With eight rounds testing up to 250 sequences each, the authors obtained higher fitness values compared to other methods, including CbAS and FBGAN. However, the authors also show that the proposed sequence diversity quickly drops, and only the diversity term added to the expected reward prevents it from converging to zero. 

While significant work has been invested in optimizing protein sequences with generative models, this direction is still in its infancy, and it is not clear which approach or framework has general advantages, particularly as many of these approaches have roots in non-biological fields. In the future, balancing increased sequence diversity against staying within each model's trusted regions of sequence space \cite{brookes2019conditioning, linder2020generative_note} or other desired protein properties will be necessary to broaden our understanding of protein sequence space. 

\section{Conclusions and Future Directions}

Machine learning has shown preliminary success in protein engineering, enabling researchers to access optimized sequences with unprecedented efficiency. These approaches allow protein engineers to efficiently sample sequence space without being limited to nature's repertoire of mutations. As we continue to explore sequence space, expanding from the sequences that nature has kindly prepared, there is hope that we will find diverse solutions for myriad problems \cite{nobeli2009protein}.

Many of the examples presented required testing many protein variants, and many of the advances in machine learning have been driven by data collection as well. For example, a large contribution to the current boom in deep learning can be traced back to ImageNet, a database of well-annotated images used for classification tasks \cite{deng2009imagenet}. For proteins, a  well-organized biannual competition for protein structure prediction known as CASP (Critical Assessment of Protein Structure Prediction) \cite{moult1995large} enabled machine learning to push the field forward \cite{senior2020improved}. A large database of protein sequences also exists \cite{uniprot2019uniprot} with reference clusters provided \cite{suzek2007uniref, suzek2015uniref}. However, these sequences are rarely coupled to \textit{fitness} measurements and if so, are collected in diverse experimental conditions. While databases like ProtaBank \cite{wang2018protabank} promise to organize data collected along with their experimental conditions, protein sequence design for diverse functions has yet to experience its ImageNet moment.

Fortunately, a wide variety of tools are being developed for collecting large amounts of data, including deep mutational scanning \cite{fowler2014deep} and methods involving continuous evolution \cite{esvelt2011system, morrison2020developing, zhong2020automated}. These techniques contain their own nuances and data artifacts that must be considered \cite{eid2021systematic}, and unifying across studies must be done carefully \cite{dunham2020exploring}. While these techniques currently apply to a subset of desired protein properties that are robustly measured, such as survival, fluorescence, and binding affinity, we must continue to develop experimental techniques if we hope to model and understand more complex traits such as enzymatic activity.


In the meantime, machine learning has enabled us to generate useful protein sequences on a variety of scales. In low- to medium-throughput settings, protein engineering guided by discriminative models enables efficient identification of improved sequences through the learned surrogate fitness function. In settings with larger amounts of data, deep generative models have various strengths and weaknesses that may be leveraged depending on design and experimental constraints. By integrating machine learning with rounds of experimentation, data-driven protein engineering promises to maximize the efforts from expensive lab work, enabling protein engineers to quickly design useful sequences. 

\section*{Acknowledgements}

The authors wish to thank members Lucas Schaus and Sabine Brinkmann-Chen for feedback on early drafts. This work is supported by the Camille and Henry Dreyfus Foundation (ML-20-194) and the NSF Division of Chemical, Bioengineering, Environmental, and Transport Systems (1937902). 

 \bibliographystyle{elsarticle-num} 
 \bibliography{citations}

\appendix

\section{Appendix: Deep Generative Models of Protein Sequence}
\label{sec:sample:appendix}
\begin{figure}
	\centering
	\includegraphics[scale=0.7]{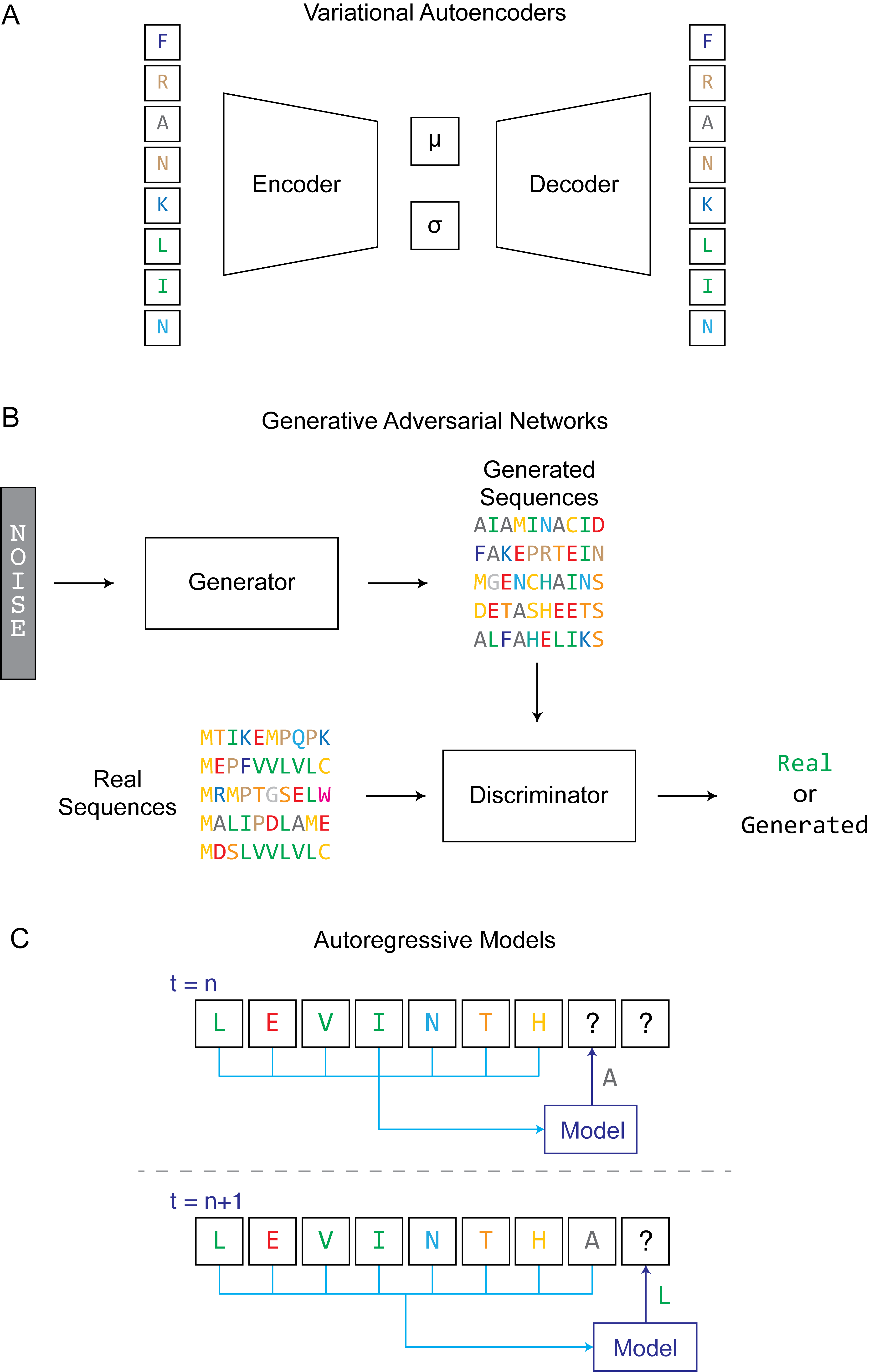}
	\caption{(A) Variational Autoencoders (VAEs) are tasked with encoding sequences in a structured latent space. Samples from this latent space may then be decoded to functional protein sequences.  (B) Generative Adversarial Networks (GANs) have two networks locked in a Nash equilibrium: the generative network generates synthetic data that look real, while the discriminative network discerns between real and synthetic data. (C) Autoregressive models predict the next amino acid in a protein given the amino-acid sequence up to that point.}
	\label{fig:rev_gen}
	\index{figures}
\end{figure}

Here, we describe three popular generative models, variational autoencoders, generative adversarial networks, and autoregressive models, and provide examples of their applications to protein sequences. These models are summarized in Figure \ref{fig:rev_gen}.

\subsection{Variational Autoencoders}
To provide an intuitive introduction to Variational Autoencoders, we first introduce the concept of autoencoders \cite{kingma2013auto, rezende2014stochastic, doersch2016tutorial}, which are comprised of an encoder and a decoder. The encoder, $q(z|x)$, maps each input $x_i$ into a latent representation $z_i$. This latent representation is comparatively low dimension to the initial encoding, creating an information bottleneck that forces the autoencoder to learn a useful representation. The decoder, $p(x|z)$, reconstructs each input $x_i$ from its latent representation $z_i$. During training, the goal of the model is to maximize the probability of the data $p(x)$, which can be determined by marginalizing over $z$:

\begin{equation}
    p(x) = \int p(x|z) p(z) dz
    \label{eqn:vae-likelihood}
\end{equation}

$p(z)$ is the prior over $z$, which is usually taken to be $\text{normal}(0, 1)$. Direct evaluation of the integral in Equation~\ref{eqn:vae-likelihood} is intractable and is instead bounded using variational inference. It can be shown that a lower bound of $p(x)$ can be written as the following \cite{kingma2013auto}:

\begin{equation}
    \log p(x) \ge \mathbb{E} \left[ \log p(x|z) \right] - \mathbb{D}_{KL} \left[ q(z|x) || p(z) \right]
\end{equation}

where $\mathbb{D}_{KL}$ is the Kullback-Leibler divergence, which can be interpreted as a regularization term that measures the amount of lost information when using $q$ to represent $p$, and the first expectation $\mathbb{E}$ term represents reconstruction accuracy. VAEs are trained to maximize this lower bound on $\log p(x)$, thus learning to place high probability on the training examples. The encoder representation can be used for downstream prediction tasks, and the decoder can be used to generate new examples, which will be non-linear interpolations of the training examples. Intuitively, the prior over $z$ enables smooth interpolation between points in the latent representation, enforcing structure in an otherwise arbitrary representation.

\subsection{Generative Adversarial Networks}
Generative Adversarial Networks (GANs) are comprised of a generator network $G$ that maps from random noise to examples in the data space and an adversarial discriminator $D$ that learns to discriminate between generated and real examples \cite{goodfellow2014generative}. As the generator learns to generate examples that are increasingly similar to real examples, the discriminator must also learn to distinguish between them. This equilibrium can be written as a minimax game between the Generator $G$ and Discriminator $D$, where the loss function is:

\begin{equation}
    \underset{G}{\min} \underset{D}{\max} L(D,G) = \mathbb{E}_{x \sim p_{real} (x)}[\log D(x)] + \mathbb{E}_{z \sim p(z)}[\log (1-D(G(z)))] 
\end{equation}

where the discriminator is trained to maximize the probability $D(x)$ when $x$ comes from a distribution of real data, and minimize the probability that the data point is real ($D(G(z))$) when the data is generated ($G(z)$). GANS do not perform representation learning or density estimation, but on image data they usually generate more realistic examples than VAEs \cite{theis2015note, dumoulin2016adversarially}. However, the Nash equilibrium between the generator and discriminator networks can be notoriously difficult to obtain in practice \cite{salimans2016improved, mescheder2018training}.

\subsection{Autoregressive models}
An emerging class of models from language processing has developed from self-supervised learning of sequences. After masking portions of sequences, deep neural networks are tasked with generating the masked portions correctly, as conditioned on the unmasked regions. In the autoregressive setting, models are tasked with generating subsequent tokens based on previously generated tokens. The probability of a sequence can then be factorized as a product of conditional distributions:

\begin{equation}
    p(\mathbf{x}) = \prod_{i=1}^{N} p(x_i|x_1, ..., x_{i-1})
\end{equation}

Alternately, the masked language model paradigm takes examples where some sequence elements have been replaced by a special mask token and learns to reconstruct the original sequence by predicting the identity of the masked tokens conditioned on the rest of the sequence: 

\begin{equation}
    p(\mathbf{x}) = \prod_{i \in \text{masked}} p(x_i|x_{j \neq i})
\end{equation}

Autoregressive models learn by maximizing the probability of the training sequences. They can be used to generate new sequences, and depending on the architecture, they can usually provide a learned contextual representation for every position in a sequence. While masked language models are not strictly autoregressive, they often use the same model architectures as autoregressive generative models, and so we include them here. 

The main challenge is in capturing long-range dependencies. Three popular architectures, dilated convolution networks, recurrent neural networks (RNNs), and Transformer-based models, take different approaches. Dilated convolution networks include convolutions with defined gaps in the filters in order to capture information across larger distances \cite{yu2015multi, oord2016wavenet}. RNNs attempt to capture positional information directly in the model state \cite{mikolov2010recurrent, kalchbrenner2013recurrent}, and an added memory layer is introduced in Long Short-Term Memory (LSTM) networks to account for long-range interactions  \cite{hochreiter1997long, sutskever2014sequence, cho2014learning}. Finally, Transformer networks are based on the attention mechanism, which computes a soft probability contribution over all positions in the sequence \cite{bahdanau2014neural, luong2015effective}. They were also developed for language modeling to capture all possible pairwise interactions \cite{vaswani2017attention, devlin2018bert, radford2019language, wolf2019huggingface}. Notably, Transformer networks are also used for autoencoding pretraining, where tokens throughout the sequence (regardless of order) are masked and reconstructed \cite{devlin2018bert}.





\end{document}